\documentstyle[twoside,fleqn,espcrc2,epsfig]{article}
\def\etal{{\it et.al.}\/}

\def\eg{{\it e.g.}\/}

\def\Fiissm{\Phi_{i,SSM}}
\def\Fii{\Phi_i}

\def\FiBe{\Phi_{Be}}
\def\FiB{\Phi_{B}}

\def\permille{$^\circ/_{\circ \circ} $}

\def\ug{$U$}
\def\N_F{$n_F$}
\def\nnu{\nu}
\def\oomega{\delta \nu}

\def\wDelta{\delta}
\def\Sppzero{S_{pp}}

\def\lapprox{\mathrel{\mathop
  {\hbox{\lower0.5ex\hbox{$\sim$}\kern-0.8em\lower-0.7ex\hbox{$<$}}}}}
\def\gapprox{\mathrel{\mathop
  {\hbox{\lower0.5ex\hbox{$\sim$}\kern-0.8em\lower-0.7ex\hbox{$>$}}}}}
\def\permille{$^\circ/_{\circ \circ}$ }

\title{HELIOSEISMOLOGY, SOLAR MODELS AND  NEUTRINO FLUXES}
\author{    
    V. Castellani\address{Dipartimento di Fisica dell'Universit\`a di Pisa,
       piazza Torricelli 1, I-56100 Pisa, Italy\\
       and Istituto Nazionale di Fisica Nucleare, LNGS,
         I-67010 Assergi (L'Aquila),Italy}, 
       S. Degl'Innocenti\address{Dipartimento di Fisica dell'Universit\`a di Pisa,
       piazza Torricelli 1, I-56100 Pisa, Italy\\
       and Istituto Nazionale di Fisica Nucleare, Sezione di Ferrara, 
       Via Paradiso 12, I-44100 Ferrara, Italy},
       W.A. Dziembowski\address{Copernicus Astronomical Center, 
       ul. Bartycka 18, P-00716 Warsaw, Poland},
       G. Fiorentini\address{Dipartimento di Fisica 
         dell'Universit\`a di Ferrara, I-44100 Ferrara \\
        and Istituto Nazionale di Fisica Nucleare, Sezione di Ferrara,
        Via Paradiso 12, I-44100 Ferrara, Italy}
        and B. Ricci\address{Istituto Nazionale di Fisica Nucleare, Sezione di Ferrara,
        Via Pardiso12, I-44100 Ferrara, Italy}
        }

\begin{document}

\begin{abstract}
We present our results concerning
a systematical analysis of helioseismic implications on solar 
structure and neutrino production. We find 
Y$_{ph}=0.238-0.259$, $R_b/R_\odot=0.708-0.714$ and
$\rho_b=(0.185-0.199)$ 
gr/cm$^3$. In the interval $0.2<R/R_\odot<0.65$, the quantity
$U=P/\rho$ is determined with and accuracy of $\pm 5$\permille~or better.
At the solar center still one has remarkable accuracy, $\Delta U/U <4\%$.
We compare the predictions of recent solar models
(standard and non-standard)  with the helioseismic results.
By constructing helioseismically constrained solar models, the central solar
temperature is found to be $T=1.58 \times 10^7$K with 
 a conservatively estimated  accuracy of 1.4\%, 
so that the major unceratainty on neutrino fluxes
is due to nuclear cross section and not to solar inputs.
\end{abstract}

\maketitle

\section{Introduction}
Helioseismology allows us to look into the deep interior of the Sun,
probably more efficiently than with neutrinos. The highly precise 
measurements of frequencies and the tremendous number of 
measured lines  enable us to extract the values of sound 
speed inside the sun with accuracy better than 1\%.  Recently it 
was demonstrated that a comparable accuracy can be obtained for 
the inner core of the Sun \cite{eliosnoi}.

The present work summarizes the results of our group in the last
year concernig a systematic analysis of helioseismic implications
on solar structure and neutrino production.

We quantitatively estimated  the  accuracy
of solar structure properties
as inferred from the measured frequencies
through the so called inversion method. This analysis provided 
the base for quantitative tests of solar models. These
tests are briefly presented  here, see  \cite{eliosnoi,eliospp,eliosmix}
for more details. 

Concerning the organization of the paper, after a short introduction to
helioseismic data and their interpretation (sect. \ref{secw}), we shall discuss 
the accuracy of solar properties as deduced  from helioseismology 
(sect. \ref{sech}).

Next (sect. \ref{secSSM})
 we shall confront with helioseismology the predictions of Standard
Solar Models (SSMs). These tests are really a big success of recent SSMs, 
{\em including} elements diffusion, 
so that one can gain further confidence in the
predicted central solar temperature and $T_c$ and neutrino fluxes $\Fii$.

However, one can take a somehow different attitude. The richness of
helioseismic data can be used so as to supply information on some of the 
less certain  inputs of solar models (\eg~solar opacity). This is
the approach we shall pursue to build Heliosesimically Constrained Solar Models
(HCSMs), to be discussed in sect. \ref{secHCSM}.

From the analysis of HCSMs it comes out that $T_c$
can be predicted with an accuracy of about one percent, a result 
obtained  even if some of
the most controversial assumptions of SSM are released. 

This result contradicts recently advocated non-standard 
solar models, build so as to significantly decrease $T_c$.
Actually we  show explicitely that these models are in
conflict with helioseismic data, see sects. \ref{secmix} and \ref{secspp}.

Future prospects and applications of helioseismology are presented
in sect. \ref{future}.

\section{Helioseismic data and their interpretation }
\label{secw}

Traditional and still most important helioseismic observables 
are frequencies of normal modes of solar oscillations. This is unlike  
in geophysical seismic sounding where the primary data are travel times of 
seismic waves. 
A possibility of measuring the wave travel times in the Sun has been discovered 
not long ago \cite{refn1}. The travel time data have been
successfully applied to probing subsurface flows 
\cite{refn2,refn3}. The possibility of sounding  
short lived phenomena localized in the subsurface layers is the 
advantage over the sounding based on 
the frequency data. Here, however, we are interested in sounding 
mean and solar structure and therefore we will not discuss further this new 
method.  

The frequencies of solar oscillations are deduced from time series 
of the intensity or radial velocity data. The latter are much more 
frequently used. Long-time continuous monitoring is essential 
for precise frequency determination. Four ground-based networks of 
automatic telescopes devoted to helioseismic observations are 
currently operated. The BISON \cite{refn4} and IRIS \cite{refn5}
 networks measure radial 
velocity variations for the unresolved solar images. With this method,
only modes of low spherical harmonic degrees ($\ell\le 3$)are detectable.
The GONG \cite{refn6}
 and TON \cite{refn7} networks use imaging instruments allowing to
detect modes up to $\ell\approx 250$ and $\ell\approx 700$, respectively.
There are three seismic instruments on the board of the SOHO spacecraft 
which started operation at the beginning of 1996. The GOLF \cite{refn8} 
instrument is 
similar to that used by IRIS. The LOI instrument \cite{refn9}, which measures 
intensity variation, has a capability for detecting modes with $\ell\le 8$. 
The third is MDI, which an imaging instrument \cite{refn10}. In the 
continuous mode of operation it detects modes up to
$\ell=250$.

Data for seismic probing from the imaging instruments are usually provided 
in the form of centroid frequencies, $\nu_{\ell n}$, and splitting 
coefficients, $a_{k\ell n}$, for the ${\ell,n}$ multiplets, where 
$n$ denotes radial order of the mode. The coefficients  
describe the azimuthal dependence in oscillation frequencies according
to the following relation 
$$\nu_{\ell nm}=\nu_{\ell n0}+\sum_{k=1}a_{k\ell n}{\cal P}_k(m),$$
where subscripts $n$ and $m$ denote
radial and azimuthal orders, respectively, and
functions ${\cal P}_k(m)$ are $k$-order orthogonal polynomials. Data from 
the GONG network and from the MDI instrument provide values of $\nu_{ln0}$ 
and six $a_k$ coefficients for over
2000 $(ln)$ multiplets. For smaller number of multiplets the coefficients 
up to $k=36$ are available in the data sets from the MDI instrument. 
 
Values of $a_k$'s
reveal that the interior rotation rate is not very different from the
surface rate and that  
the Sun is a very slowly rotating and almost 
perfectly spherical star. This fact significantly simplifies 
interpretation of the data. In particular, the
quantities  $\nu_{ln0}$, $a_{2k+1,ln}$, and  $a_{2k,ln}$ 
are separate probes of, respectively, the radial structure, the differential 
rotation and the asphericity. The possibility of sounding internal rotation 
is of great importance for understanding mechanism of the solar magnetic 
activity changing in the 11-year cycle. The data reveal that the differential
rotation observed in the surface persists through the convective zone
(some outer 30\% in the radius). The transition to essentially uniform 
rotation takes place in the narrow zone, which has been the suggested site 
of the solar dynamo \cite{refn11}. An important result in the context of 
this review is that there is no evidence for the rotation 
increase toward the center. In fact, according to some studies 
(e.g. \cite{refn12}) there is a 
seismic evidence 
for lower rotation rate in the core than in outer layers. In any case 
the dynamical effect of centrifugal force is certainly negligible.
There is also no evidence for a significant role of magnetic field 
in the Sun's interior. 
 

\begin{figure} [tb]
\vspace{-1.3cm}
\epsfig{file=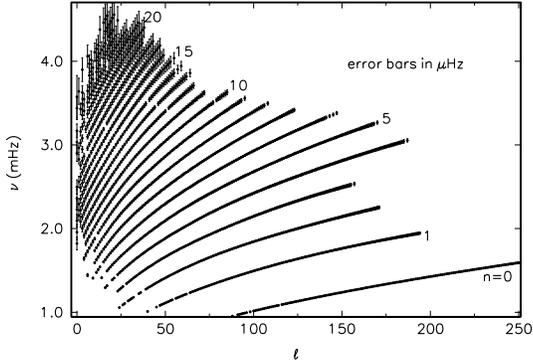,
width=0.9\hsize}
\caption{Frequencies of solar oscillations determined from 360-day measurements 
with the MDI/SOHO instrument. The error bars 
have been magnified by factor 1000.}
\label{fig1w}
\end{figure}

We now focus on the centroid frequencies and, as an example,
on the data from the 
MDI/SOHO instrument. The frequencies with 
the error bars are shown in Fig. \ref{fig1w}. The accuracy, which at the low 
frequencies is $\sim 10^{-5}$, visibly deteriorates at higher frequencies.
The peaks in this part of the oscillation amplitude spectra are 
considerably broader,
which is connected with the shorter life-time of the higher frequency modes.
The total number of frequency data in the figure is 2047. The  
$n=0$ branch represents fundamental modes which are horizontally propagating
waves obeying the same dispersion as surface waves in deep water. Their 
frequencies are fully determined by mean density in the Sun. 
The $n\ge 1$ branches represent p-modes which are standing 
acoustic (pressure) waves. These waves begin their downward propagation
from the surface, where they are reflected, as nearly vertical 
rays. As a result of increasing adiabatic sound speed ($a$), the rays  become  
gradually more oblique and, eventually, are reflected at the surface where
the sound speed  satisfies the condition 
$a(R)/R=2\pi\nu/\sqrt{\ell(\ell+1)}$.

Only a small fraction of p-modes from Fig. \ref{fig1w} penetrates the energy 
producing core. At $\nu\approx3$ mHz and $\ell=1$, the inner reflection 
surface occurs at $R=0.06R_\odot$, which is close to the maximum 
the differential $^8$B-neutrino flux of and well beneath the maximum 
of the differential photon flux. However, at a similar frequency and 
$\ell=5$ the reflection takes place at $R=0.18R_\odot$, where the photon flux 
is already more than 92\% of the total solar luminosity.
Even for the low degree modes the frequencies are 
mainly sensitive to the structure of the outer layers. The modes which 
are very sensitive to the core structure are g-modes. 
Their frequencies, however, are below 
0.43 mHz and we may only hope that the solar g-modes will be some 
times detected.


\begin{figure} [tb]
\vspace{-1.3cm}
\epsfig{file=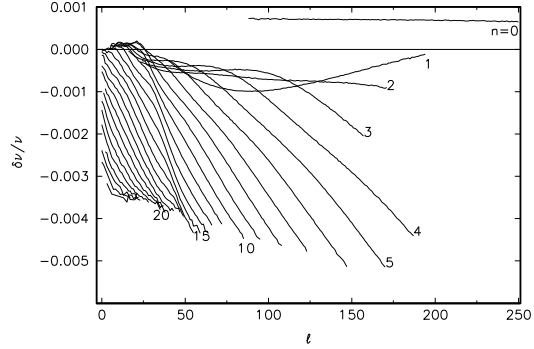,
width=0.9\hsize}
\caption{
Relative differences between 
solar (shown in Fig. \ref{fig1w}) and frequencies calculated
for standard solar model,
``model S'' of Ref. [30] .             }
\label{fig2w}
\end{figure}

In Fig.  \ref{fig2w} we show the relative differences in frequency between 
the Sun and one of the standard solar models. 
Similar patterns are found for other models.
The differences  
are small in the sense that there is no ambiguity in assigning 
the $n$ value. However, they exceed the measurement 
errors frequently by two orders of magnitude. Thus, they  
are very significant and must be explained. 
In the case of the f-modes 
almost all the difference may be removed by an adjustment of the model radius 
\cite{refn13}. 
Solar seismic radius is by 0.045\% less than adopted in the model.  
For higher order p-modes the difference between model and 
solar frequencies rapidly increases with $\ell$. This immediately suggests 
that most of the difference between the model and the Sun must be localized in
outermost layers. Higher $\ell$ implies a higher $R$ at the lower reflecting 
surface and therefore a greater confinement to outer layers. 

We know that our treatment of the structure and oscillations in outer 
layers is inadequate. The problem is how to take into account effects of 
vigorous and nonadiabatic convection. We expect 
the problem concerns the layers above $R=0.99R_\odot$, where p-modes 
propagate almost vertically and the local properties are $\ell$-independent.
This implies that the part of the frequency difference may be modeled in 
the form $F(\nu)/I$, where $I$ is the calculated mode inertia assuming uniform 
normalization and $F$ should be determined from the data. It is found 
that for modern standard solar models this part dominates in $\wDelta \nu$ 
but we are interested only in the small contribution arising in the 
rest of the interior.

In the deeper layers the neglect of dynamical effects of convection is fully 
justified. 
Furthermore, the nonadiabatic effects in oscillations, which have not 
been taken into account in calculation of frequencies used in Fig. \ref{fig2w}, 
are certainly 
negligible below $R=0.99R_\odot$. In view of the small values of 
$\wDelta\nu/\nu$, linearization about the reference standard 
model seems justified. Thus, 
we may use the variational principle for adiabatic stellar oscillations
to connect the frequency difference to the differences 
in structural functions between the Sun and its model. Our aim is 
to determine these differences. In general, 
we have to consider simultaneously two unknown 
functions of $R$. For one the choices are differences in density, $\rho$, 
pressure $P$, or
any combination of them or their derivatives. All such functions are linked 
by linearized hydrostatic condition. We use here $U=P/\rho$. 
For the other thermodynamic function the choices are the adiabatic exponent
$\Gamma_1=(\partial\ln P/\partial\ln\rho)_{\rm ad}$ or squared adiabatic 
sound speed, $a^2=U\Gamma_1$. The problem is simplified if we make use the 
$\Gamma_1(P,\rho,Y)$ relation, 
where $Y$ is the fractional helium abundance, 
from the astrophysical equation of state data.
(see e.g. Ref. \cite{refn13}). 
Since in the chemically inhomogeneous interior the 
we may safely neglect $\wDelta \Gamma_1$, the problem may be reduced to  
determination of the function $\wDelta U(R)$ and the number 
$\wDelta$Y$_{ph}$-- the 
abundance of He in the layers above the base of the convective zone.
In this way the basic equation for 
seismic probing of the internal structure becomes:
\begin{equation}
\bigl({\wDelta \nu\over \nu}\bigr)_j=\int{\cal
K}_j{\wDelta U\over U}dR+ {\cal J}_j\wDelta \mbox{Y}_{ph}
+{F(\nu)\over I_j}\, ,
\end{equation}
where $j\equiv(n,\ell)$. The kernel $K_j$ and numbers $J_j$, $I_j$ are 
easily evaluated in terms of the eigenfunctions for adiabatic 
oscillations in the standard solar model.


\begin{figure}[tb]
\vspace{-1.3cm}
\epsfig{file=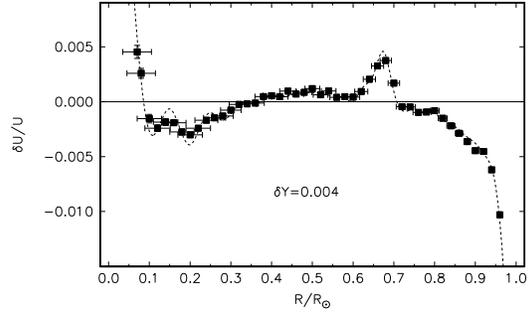,
width=0.9\hsize}
\caption{Relative differences in $U=p/\rho$ between the Sun and the 
model inferred from the differences in $\nu$ shown in Fig. \ref{fig2w}. The
symbols with error bars give the averaged values  with the Gaussian-like 
localized kernels. The horizontal error bars give the full width
at half-maxima of the kernels. Vertical error bars directly reflect the 
observational errors. The dashed line represents the result of inversion 
by means of the regularized least square method in which $\wDelta U/U$ is 
searched
in the form of a superposition of cubic splines. Near the center results
obtained with this method are unreliable. The value of $\wDelta$ Y$_{ph}=
(4.0\pm 0.1)\times 10^{-3} $ was 
determined simultaneously with  $\wDelta U/U$.}
\label{fig3w}
\end{figure}

This equation may be applied to all p-modes. 
Thus, we have 
a large set of the integral equations to determine the two functions 
and the Y$_{ph}$. Methods of 
determination of $\wDelta U/U$ were described in details in 
Refs. \cite{ref10}
and \cite{rhow}.
There are two distinct approaches. The first is the regularized 
least squares (RLS) method 
 which consists in a discretization in terms of known function and 
a determination of parameters by means of minimization of the $\chi$-squared 
fit plus an integral term that smooths artificial oscillation.
The other is the optimal averaging method which consists in seeking linear 
combination of individual kernels which are close to possibly narrow
Gaussians centered at specified distances $R$. Corresponding 
combination of frequency differences yields the averaged values of 
$\wDelta U/U$ weighted with these localized kernels. Fig. \ref{fig3w}  
shows that the results obtained with these  methods 
agree in a good agreement.

Once we know $\wDelta U(R)$ we may determine $\wDelta \rho(R)$,
$\wDelta P(R)$ and $\wDelta M_R$ making no additional assumptions.
Hence we get seismic values of these three parametrs in the 
whole solar interior. At this stage we can determine the value $R$ at
the base of the convective, $R_b$, as well as the temperature 
$T(R)$ for $R\ge R_b$. 
What we cannot find is $T(R)$  
$Y(R)$  profiles in the radiative interior. In order to disentangle 
the two last functions, which are of course most interesting in the 
context of the solar neutrino problem, we have to use data on
the opacities and  the nuclear reaction rates. This has been done 
(see Ref. \cite{ref10} for references to the original work) but we will not 
pursue this way here. Rather, we rely on directly inferred quantities as 
constraints on solar models.

We see in Fig. \ref{fig3w} that the relative differences in $U$ are 
less than $5\times 10^{-3}$ in the whole $[0.05-0.95]R_\odot$ range 
of $R$.  This agreement is very unlikely accidental and, thus, 
we regard it as
a confirmation of the standard model of the solar evolution. 
In the outer part of the convective zone 
the agreement is worse but it may be easily improved by admissible 
modifications in the description of the convective energy transfer. 
Beneath $R=0.05R_\odot$ we just do not have enough information to make 
any statement.
 The small 
 differences in the intermediate region may be eliminated 
by introducing {\it ad hoc} opacity modifications which are within 
the uncertainty of its calculation \cite{refn15}.
However, the bump in $\wDelta U/U$ near the bottom of the 
convective envelope may have another cause. 
Richard et al.  \cite{Vau} showed that this 
feature  may be removed if one allows a weak rotation-induced
mixing of elements below the bottom of the convective envelope. 
The effect was considered  
as a possible explanation of the deficiency of lithium 
in the Sun's atmosphere. Mixing brings lithium to deeper layers 
where it burns. It also results in 
a reduction efficiency of the gravitational settling leading to
somewhat higher Y$_{ph}$ in the convective zone. Thus, the 
same effect may 
also explain the seismic 
correction to the photospheric He abundance 
($\wDelta$ Y$_{ph}=4\times 10^{-3}$).

\begin{table*}[htb]
\setlength{\tabcolsep}{1.4pc}
\newlength{\digitwidth} \settowidth{\digitwidth}{\rm 0}
\catcode`?=\active \def?{\kern\digitwidth}
\caption[errori]{
For the indicated quantities $Q$ we present the helioseismic
values
$Q_\odot$ and the relative  errors $\Delta Q/Q$.
 All uncertainties are in  \permille. 
In the  fifth and sixth row, 
for  $U=P/\rho$ the values of 
 the uncertainties are the maxima in the 
indicated interval.
In the last two rows the results on $U$ at points representative of the 
$^7$Be  and $^8$B neutrino production are shown.
               }
\label{tabq}
\vspace{0.2cm}
\begin{tabular}{lcccccccc}
\hline
$Q$ &&&$Q_\odot$&&&&$\left ( \frac{\Delta Q}{Q}  \right )  $& \\
\hline
Y$_{ph}$ &&&0.249&& &&  42& \\
$R_b/R_\odot$& &&0.711&&&& 4 & \\
$\rho_b$ [g/cm$^3$] &&& 0.192 &&&& 37  & \\
\hline
$U$($0.2<x<0.65$ )&&& &&&& 5& \\
$U$($0.1<x<0.2$ )&&&  &&&& 9.4&  \\
\hline
$U$$(x_{Be})$ [$10^{15}$ cm$^2$ s$^{-2}$]&&& 1.56  &&&& 17  &\\
$U$$(x_{B})$  [$10^{15}$ cm$^2$ s$^{-2}$] &&& 1.56 &&&&  22 &\\
\hline
\end{tabular}
\end{table*}

\section{How accurate are solar properties as inferred 
from helioseismology?}
\label{sech}

We performed 
a systematic and possibly exhaustive investigation of the uncertainties of the 
helioseismic approach, in order to estimate the global error to be assigned to 
helioseismic determinations of  solar properties.
 With this spirit, we  analyse  
several physical quantities $Q$ characterizing the solar structure.
Concerning the outer part of the sun, we  discuss  the photospheric
helium abundance
Y$_{ph}$,  the depth of the the convective envelope $R_b$, 
and the density  at the 
bottom of the convective zone  $\rho_b$.
Then  we  consider  the ``intermediate" solar interior ($x$=$R/R_\odot =0.2-
0.65$), analysing the behaviour of the squared isothermal sound speed,
\ug=$P/\rho$.
Finally we  investigate the inner region ($x \leq 0.2$), where 
nuclear energy  and neutrinos are produced.  

We remind --see the previous section--
that helioseismology  measures {\em  only } the frequencies  
\{$\nnu$\} of solar p-modes, and  quantities characterizing the solar 
structure are indirectly inferred from the \{$\nnu$\}'s, through an inversion 
method.  Schematically, the procedure is the following:

a)One starts with a solar model, giving values $Q_{mod}$ 
 and predicting a set  \{$\nnu_{mod}$\} of frequencies.  
These  will be somehow different from the measured frequencies, 
$\nnu_\odot \pm \Delta \nnu_\odot$

b)One  then searches for  the corrections  $\wDelta Q$ to the solar model which 
are needed in order to match the corresponding frequencies
 \{$\nnu_{mod} + \oomega$\}  with the observed 
frequencies \{$\nnu_{\odot}$\}. 
Expression for $\oomega$ are derived 
 by using 
perturbation theory, where  the starting model is used as a zero-th order 
approximation. 
The correction factors $\wDelta Q$ are then computed, assuming some regularity
 properties, so that
 the problem is  mathematically well defined and/or 
unphysical solutions are avoided.

c)The ``helioseismic value'' $Q_\odot$ is thus determined by adding 
the starting value and the correction 
\footnote{
Concerning notation, we remark that $\wDelta Q$ indicates the correction to
solar model to obtain helioseimic value (see Eq. \ref{corr}),
 whereas $\Delta Q$ indicates
the estimated uncertainty on $Q$.}:
\begin{equation}
\label{corr}
Q_\odot= Q_{mod} + \wDelta Q  \, .
\end{equation}

For each quantity $Q$  we have determined the partial errors 
corresponding 
to each  uncertainties of the helioseismic method.
In fact, there are three independent sources of errors in the 
inversion process:

i)Errors on the measured frequencies, which -- for a given inversion 
procedure -- propagate on  the value of $Q_\odot$.

ii)Residual dependence on the starting model:
the resulting $Q_\odot$ is slightly different
if one starts with  different   solar models. This introduces an 
additional uncertainty, which can be evaluated by 
comparing the results of several calculations.

iii)Uncertainty in the regularization procedure. Essentially this is a 
problem of 
extrapolation/parametrization. Different methods, equally acceptable in 
principle, yield (slightly) different values of $Q_\odot$.

It has to be remarked that, in view of the extreme precision of the measured 
frequencies, $ \Delta \nnu_{\odot} / \nnu_{\odot} \lapprox 10^{-4}$ 
\cite{ref1,ref2,ref3,ref4}, 
uncertainties corresponding to ii) and iii) are extremely important.

For deriving a global uncertainty, we took a very
conservative approach.  May be that the parameter variation was not 
exhaustive, and what we found as extrema are not really so, but actually are 
quite acceptable values. In view of this, let us double  the interval we found 
and interpret $\pm( \Delta Q  )_k$, as partial errors. Furthermore, let us be 
really {\em conservative} 
assuming that errors add up linearly. In conclusion, this gives:
\begin{equation}
\label{cons}
\Delta Q =  \pm  \sum_k  | (\Delta Q)_k |  \, .
\end{equation}

In the following sections, we shall use this error estimate.

\subsection{Properties of the convective envelope}

Three independent physical 
properties of the convective envelope 
 are determined most accuratelly by seismic observations, 
see Ref.\cite{eliosnoi}:
Y$_{ph}$, $R_b$ and
$\rho_b$
\footnote{A fourth seismic ``observable", the sound speed at the convective radius
is traditionally considered, e.g. \cite{c-d}. We have not included it in 
our list 
since, as shown in Ref.\cite{elios}, it is not an independent one.}.

The helioseismic predictions, together with their (conservatively estimated)
accuracy, are shown in Table \ref{tabq}. 
We remark that for all these quantities 
the uncertainty resulting from propagation 
of the frequency measurement errors
is of minor importance with respect to the 
``systematic'' errors, intrinsic to the 
inversion method.

\subsection{The intermediate region}

The essential output of helioseismology is the reconstruction 
of the adiabatic sound speed profile, $a$. Our discussion is in terms of the
related quantity $U$; as well known (see also sect. \ref{secw}),
 $a^2$=$\Gamma_1 U$ 
and $\Gamma _1$ is given by the equation of state with an
accuracy of 10$^{-3}$ or better (even the simplest EOS, fully
ionized perfect classical gas yields 
$\Gamma _1$ = 5/3 with an accuracy of about 10$^{-3}$). 

By using the RLS method (see sect. \ref{secw})
 it is possible to derive  directly the profile 
of  $U$  as a function of the 
radial coordinate throughout all the sun, except for  the inner region
 ($x<0.1$).
The results (values of $U$
 and global errors), are summarized in Table \ref{tabq}.

\begin{figure}[tb]
\vspace{-1.3cm}
\epsfig{file=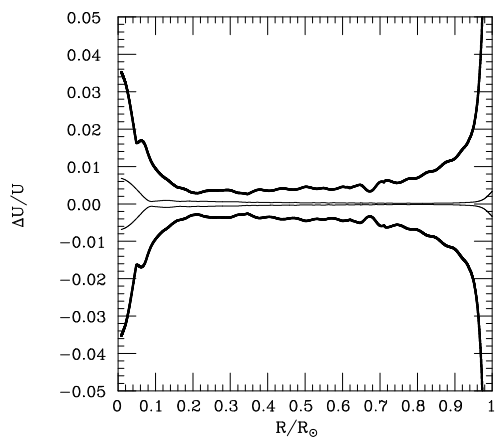,
width=0.9\hsize,angle=90}
\vspace{-2.1cm}
\caption[aaa]{
The estimated global relative uncertainty on $U=P/\rho$
(thick line) and that due to
the observational errors (thin line).
             }
 \label{fig1}
\end{figure}

It is convenient to consider an intermediate solar region:
$0.2<x<0.65$. The upper limit is established by 
requiring that it is well below the transition to the convective zone, 
which we discussed
above. The lower limit is chosen so as to exclude the region of 
energy production 
(see next subsection).
For this  region the  following 
comments are relevant:

a)Each of the individual uncertainties nowhere exceeds 
2~\permille.  

b)Uncertainties from the accuracy on the measured frequencies 
are of minor relevance with respect to the residual model dependence and to
the sensitivity to the inversion parameters, see Fig. \ref{fig1}.

c)All in all, even with the most conservative estimates, the helioseismic
determination is extremely accurate:
$|\Delta U/U | \leq  5{\mbox{\permille}}$
throughout the explored region.


\subsection{The energy production region} 

As well known, most of the energy and of solar neutrinos originate 
from the innermost part  of the sun. According to SSM calculations,
see e.g. Refs. \cite{Report,BP95}, about  
94\% of the solar luminosity  and 93\% of the pp neutrinos are produced 
within $x<0.2$, the region which we analyse in this section. 

Our results are summarized in Fig. \ref{fig1} and  Table \ref{tabq}.
Clearly the precision worsens in this region, due to the fact
 that p-modes do not penetrate 
in the solar core, and consequently the information one can
 extract from available experimental results is limited,
but still important.

For example, at the production maxima of $^7$Be and $^8$B neutrinos
($x_{Be}=0.06$  and $x_{B}=0.04$  according to our best solar model with diffusion 
\cite{Ciacio} ) the global  accuracy is still 
a 2\%, see Table \ref{tabq}. Even at $x=0$ the accuracy is 3.5\%.
 In conclusion,
{\em helioseismology provides significant insight even on the solar
innermost core}.


\begin{figure}[tb]
\vspace{-1.3cm}
\epsfig{file=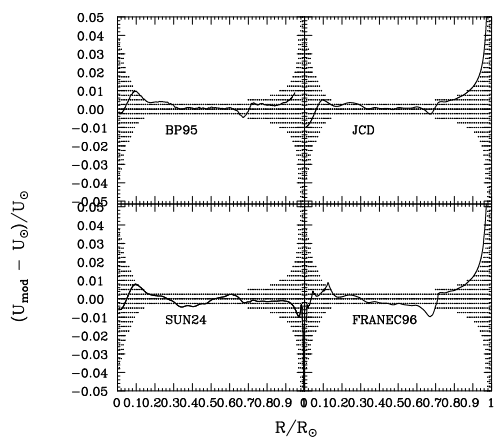,
width=0.9\hsize,angle=90}
\vspace{-2.1cm}
\caption[aaa]{
The difference between $U$ as predicted by selected solar 
models, $U_{mod}$, and the helioseismic determination, $U_{\odot}$,
 normalized to this latter. 
The dotted area corresponding to 
$\left ( \frac{\Delta U}{U}  \right ) $.
SUN24 is the ``model 0'' of Ref. 
\cite{ref10};
FRANEC96 is the ``best'' model 
with  He and heavier elements diffusion of
Ref. \cite{Ciacio};  BP95 is the model with metal and He diffusion of 
Ref. \cite{BP95}; JCD is the 
``model S'' of Ref. \cite{JCD}.             }
 \label{fig2}
\end{figure}

\section{Helioseismology and SSMs}
\label{secSSM}

\begin{figure}[tb]
\vspace{-1.3cm}
\epsfig{file=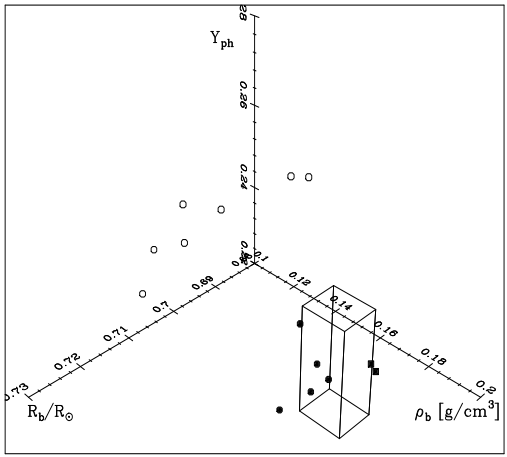,height=8cm,%
width=0.9\hsize,angle=90}
\vspace{-2.1cm}
\caption[aaa]{
Helioseismic determinations and solar model predictions of
properties of the convective envelope. 
The box defines the region allowed by helioseismology.
Open circles denote models without diffusion, squares models with He diffusion,
full circles models with He and heavier elements diffusion, see Ref. 
\cite{eliosnoi}.
             }
\label{fig3}
\end{figure}

The comparison between the predictions of a few recent SSM
calculations and helioseismic information is shown in Figs. \ref{fig2}
and \ref{fig3}.

Concerning the (isothermal) sound speed profile, see Fig. \ref{fig2},
all models look generally
good. Also SUN24, a model which neglects 
elemental diffusion, passes this test.

The study of  convective envelope is illuminating, see Fig. \ref{fig3}.
 All models neglecting 
elemental
diffusion are in clear contradiction with helioseismic constraint. On the other
hand, calculations where diffusion is included look in  substantial agreement
with helioseismology.

All this shows that the two approaches  (profile of $U$ and properties of the 
convective envelope) are complementary and both important.

The previous arguments show that SSMs are in good shape. 
Actually, helioseismology provides a  new perspective/definition of SSMs.

Before the advent of helioseismology a SSM had three essentially free parameters,
$\alpha$, Y$_{in}$ and (Z/X)$_{in}$ for producing three measured
quantities: the present radius, luminosity and heavy element 
content of the photosphere. This may not look as a too big accomplishment,
in itself.

Nowadays, by using the same number of parameters a SSM has to 
reprooduce many additional data, such as 
Y$_{ph}$, $R_b$, $\rho_b$, $U(R)$, provided
by helioseismology.

Alternative solar models have to be 
confronted with these data too.


\section{Helioseismically constrained solar models}
\label{secHCSM}


\begin{table*}[htb]
\setlength{\tabcolsep}{1.4pc}
\catcode`?=\active \def?{\kern\digitwidth}
\caption[flussi]{
Predictions of neutrino fluxes and signals in the Cl and Ga detectors 
from  HCSMs. Uncertainties corresponding
 to $(\Delta T/T )_{HCSM}=\pm 1.4\%$ 
are shown (first error) together with those from
 nuclear cross sections (second error).
               }
\label{tabsig}
\vspace{0.2cm}
\begin{tabular} {cllcccccc}
\hline
&$\FiBe$ & $[10^9$/cm$^{2}$/s$]$  &  &  &  & &  4.81$\pm  0.53 \pm 0.59$ & \\
& $\FiB$ & $[10^6$/cm$^{2}$/s$]$  &  &  &  &  & 5.96$\pm 1.49 \pm 1.93$ & \\
&Cl & $[$SNU$]$&   &  &  & & 8.4$\pm 1.9 \pm 2.2$  & \\
&Ga & $[$SNU$]$&  & & & &  133$\pm 11 \pm 8$ & \\
\hline
\end{tabular}
\end{table*}

Helioseismic data are thus in agreement with 
recent SSM calculations,
which use accurate equations of state,
recent opacity tables and include helium and heavier elements diffusion
\cite{BPelios,eliosnoi,BP95,Ciacio,JCD}, see also Ref. \cite{Vau}.
These SSMs yield central 
temperatures 
$T_{c,SSM}$ which differ from each other by not more than 1\%. However the 
uncertainties in the
input parameters, mainly the opacity $\kappa$ and the heavy elements
abundance Z/X,  result in $(\Delta T_c/T_c)_{SSM} \approx 1-2\%$.

From helioseismic observations one  cannot determine 
directly temperatures of the solar interior, as  one cannot determine
the temperature of a gas from the knowledge of the sound speed unless the 
chemical composition is known. However, it is possible 
{\em to obtain the range of 
allowed values of the central temperature $T_c$, by
selecting those solar models which are consistent with 
seismic data}.

Our calculations are not model-independent, but we shall use in 
principle a wider  class of models in comparison with SSMs, which we call 
helioseismically-constrained solar models (HCSM). 
These models are based on the same  equilibrium 
and evolution equations as SSMs, but they differ in the choice of 
some input parameters.  We generate this class
of models by using the FRANEC code \cite{Ciacio}
for the SSMs and varying the input 
parameters. 
Each choice of the set of parameters gives some value of $T_c$.

We obtain the range of allowed values of $T_c$ by selecting those solar
models which are consistent with seismic data. More specifically, we shall
determine the central temperature $T_{c,HCSM}$, as that of the model
which gives the best fit to the seismic data and the uncertainties, 
$\Delta T _{c,HCSM}$,
corresponding to the range spanned by models  consistent with these
data.

We remind that the
 precise value of the temperature is governed essentially by two quantities:
 the radiative opacity $\kappa$ and the fraction  of  heavy elements Z/X.
 As well known the 
  uncertainties on $\kappa$ and Z/X are of the order of  10\%.
  Furthermore,
these uncertainties do not correspond to clear experimental
or observational errors, rather they are determined 
by judicious comparison among published values. 
As an example, the uncertainty on $\kappa$ can only
be estimated from the comparison
among recent theoretical calculations.

We allowed that both $\kappa$ and Z/X are 
{\em rescaled by  free multiplicative factors}
with respect to the value used in th SSM.
These scaling factors are then  determined
by helioseismic constraints on the convective envelope.

In this way (see Ref. \cite{eliostcnoi}), we obtained as the best 
estimate $T_{HCSM}=1.58 \times 10^7$ K, with  a conservately estimated
uncertainty $(\Delta T/T)_{HCSM}= \pm 1.4\%$, to be
compared with the uncertainty of SSM calculations  
 $(\Delta T/T)_{SSM}= \pm 2.7 \%$.

We remark the following points:\\
i)The ``best" temperatures determined by
means of helioseismology starting from
different solar models converge.\\
ii)The uncertainty $(\Delta T_c )_{HCSM}$ is (slightly) reduced with
respect to $(\Delta T_c )_{SSM}$.\\
iii)More important, the helioseismic uncertainty is
related to observational data
(whereas as noted above  $\Delta \kappa$,
and thus $(\Delta T_c )_{SSM}$, was
derived just from comparison among
theoretical calculations).

Let us come over to {\em neutrino fluxes}, $\Fii$ ($i$=pp, $^7$Be, $^8$B).
Their dependence  on the central 
temperature $T_c$ is parametrized  as:
\begin{equation}
\label{lexf}
 \Fii =\Fiissm \left  ( \frac{T_c}{T_{c,SSM}} \right ) ^{\beta _i} \, .
\end{equation}

From numerical experiments with FRANEC, we found:
$\beta _{pp}=-0.92,\,\beta _{Be}=7.9, \, \beta _{B}=18$.

We have determined neutrino fluxes by starting with different SSMs,
renormalizing their predictions to the same temperature $T_{HCSM}=1.58\cdot
10^7 \mbox{K}\,$
and to the same (updated) nuclear cross sections. The resulting fluxes and
signals, all very close to each other, have been averaged to
determine the HCSM predictions shown in Table \ref{tabsig},
where the first error corresponds to a
conservatively estimate
$(\Delta T_{c}/T_{c})_{HCSM} =1.4 \%$.
and the scond one to a $3\sigma$  error on
nuclear cross sections.

We remark that nuclear physics uncertainties look
larger than astrophysical ones.
After the successful LUNA experiment measuring
the $^3$He+$^3$He cross section \cite{luna},
 the main  uncertainties  arise now from
the astrophysical S-factor for 
$^3$He+$^4$He and p+$^7$Be reactions,
 which should be measured
more accurately.

\section{A mixed solar core?}
\label{secmix}

Several authors have proposed non-standard solar models,
where some ad-hoc mechanism is introduced, so as to reduce
the central solar temperature by a few percent, which might
alleviate (not solve) the solar neutrino puzzle.

Clearly such  a reduction of $T_c$ is in contrast
with the conclusions of the previous section.
Here we  show explicitely  that such models
are inconsistent with helioseismic information.

As an example, consider the case of a mixed solar core,
firstly advanced by
Shaviv and Salpeter in 1968 \cite{SS68} and recently
revived by Haxton and Cumming \cite{CH97},  see
also Refs. \cite{BP95,Vau2}.
Roughly speaking, mixing of H and He over
an {\em appreciable} portion of the sun enriches
the innermost solar core with H.
Hydrogen burning  becomes more efficient and the solar
luminosity is attained at lower core
temperatures.
This results in reduced $T_c$, $^7$Be+CNO  and $^8$B neutrino
fluxes, again if a significant portion ($R > 0.1 R_\odot$)
of the Sun is mixed.

Several different mixing processes (fast or slow, continuous or episodic)
can be conceived \cite{EC68,SS68,BBU68,SS71,UR83,SBP90}. While we refer to
Ref. \cite{eliosmix} for an extended discussion, we
concentrate here on the case of fast continuos mixing.

\begin{figure}[tb]
\vspace{-1.3cm}
\epsfig{file=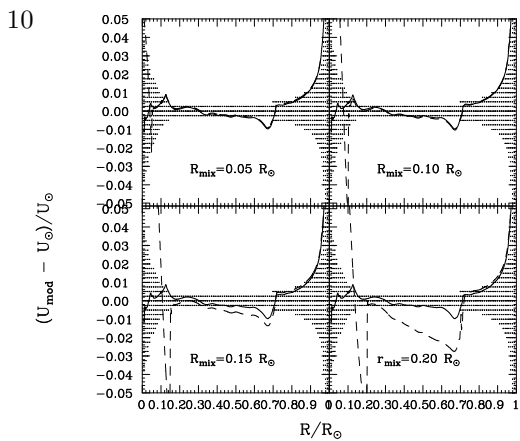,
width=0.9\hsize,angle=90}
\vspace{-2.1cm}
\caption[fast]{
For the indicated values of $R_{mix}$, we present the relative difference
between the isothermal sound speed as predicted by solar model with fast 
continuous mixing, $U_{mod}$, and the helioseismic determination,
$U_{\odot}$ (dashed lines). The same quantity for our SSM is also shown 
(solid line).
The dotted area corresponds to the uncertainty
on $U_{\odot}$.
             }
 \label{fig4}
\end{figure}

In Fig. \ref{fig4} we show (dashed line) 
 the relative difference between the isothermal sound speed
squared, $U=P/\rho$, as predicted by  the mixed models and the 
helioseismic  determination, U$_\odot$. The same quantity for
our SSM is also  shown (solid line).


One sees a strong deviation of Mixed Core Models (MCMs),
with respect to both $U_\odot$ and 
$U_{SSM}$, in the mixing zone; this is a consequence of 
the change in ``mean molecular weight'', $\mu$,  induced by mixing.
In the approximation of perfect gas  
(accurate to the level of few per thousand  in the solar core)
one has $U\propto T/ \mu$. Due to mixing, the innermost region 
is enriched with hydrogen, so that $\mu$ decreases 
(we observe that change of $\mu$ can be as high as 40\%,
whereas temperature change is at most  a few per cent) and
$U$ increases. The opposite occurs near the edge of the mixed
region.

As the mixing area increases, the sound speed profile of the
MCMs deviates more and more from the SSM prediction and
it becomes  in conflict with helioseismic constraint if $R_{mix} \ge 0.1 R_\odot$.

\begin{figure}[tb]
\vspace{-1.3cm}
\epsfig{file=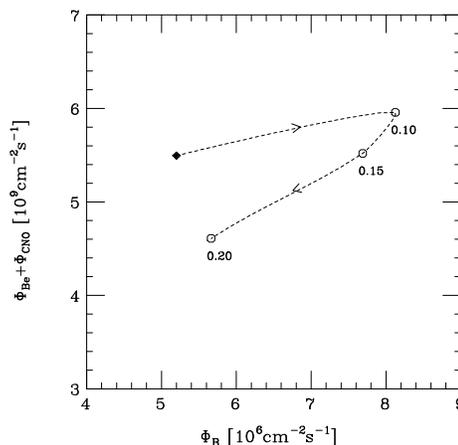,
width=0.9\hsize,angle=90}
\vspace{-2.1cm}
\caption[flussi]{
The predictions of intermediate 
($^7$Be+CNO) and high energy ($^8$B) neutrino fluxes in  solar models 
with continuous fast mixing, for the indicated values of $R_{mix}$.
The prediction of our SSM (full diamond) is also shown.
             }
 \label{fig5}
\end{figure}

Neutrino fluxes predicted by MCMs are shown in
Fig. \ref{fig5}. One sees that reduction of
intermediate ($^7$Be+CNO) and/or high ($^8$B) energy
neutrino fluxes is only achieved  for $R_{mix} \ge 0.1 R_\odot$.

Similar conclusions hold for the other mixing mechanisms mentioned
above, see Ref. \cite{eliosmix}.

In brief, {\em  reduction of $^7$Be and $^8$B neutrino fluxes
can only be obtained for extended mixing regions }
($R_{mix} \ge 0.1 R_\odot$), {\em corresponding to solar models
inconsistent with helioseismic constraints}.

We remind that the one percent accuracy of helioseismic information
at $R/R_\odot \approx 0.1$ is an essential ingredient
for achieving this result.


\section{Helioseismology and physics of fundamental interactions}
\label{secspp}

As another example illustrating the potential of the helioseismic
approach we
discuss the  p+p$\rightarrow$d+e$^+$+$\nu _e$
reaction.

The rate of the initial reaction in the pp chain is too low to be
directly  measured in
the laboratory (even  in the solar  center this rate is extremely small,
clearly of the order of 10$^{-10}$ yr$^{-1}$)
and it can be determined only by using the theory
of low energy weak interactions, together with the measured
properties of the deuteron and of the proton-proton scattering.
While we refer to Refs. \cite{BP92,KamB,Report} for  updated reviews,
we remind that, as input of SSM calculations, one
takes \cite{KamB} 
$S _ {pp,SSM} =3.89 \cdot 10^{-25} (1 \pm 0.01) \, {\mbox{ MeV b}}$.

We remark however that 
only theoretical estimates
of $\Sppzero$ are available and observational
information would be welcome. In this respect, it is interesting to
determine the range of $S_{pp}$ which is  acceptable in 
comparison with helioseismology.

In sect. \ref{secHCSM} we pointed
 that there are two major uncertainties in building SSMs:
solar opacity $\kappa$  and heavy element abundance $\zeta =$Z/X
are only known with an accuracy of about $\pm$ 10\%.
By using  now $\kappa $, $\zeta$ and as $\Sppzero$ free parameters,
we can determine the acceptable range of
$\Sppzero$ by requiring that $R_b,\, \rho_b$ and Y$_{ph}$ are
all predicted within the helioseismic range.
The dependence of these quantities on $\kappa$, $\zeta$ and $S_{pp}$
has been determined numerically in Ref. \cite{eliostcnoi}.

Most of the information on $\Sppzero$ arises from data on $\rho_b$
as this observable depends strongly on $S_{pp}$ whereas it is
weakly affected by the others parameters.
One can understand the dependence on $\Sppzero$, at least qualitatively.
A value of $\Sppzero$ larger than $S_{pp,SSM}$ 
implies smaller temperatures in the
solar interior, which thus becomes more opaque
 (in other words, the region of partial ionization is deeper).
 Radiative transport therefore
is less efficient and convection starts deeper
in the Sun ($R_b < R_{b,SSM}$),
 where density is higher ($\rho_b >\rho_{b,SSM}$).

By restricting to the allowed ranges for the properties of the
convective envelope, see Table \ref{tabq},
also taking into account the predictions of different SSMs, we find:
\begin{equation}
0.94 \leq  S/S_{SSM} \leq 1.18
\end{equation}

In conclusion, we remark that helioseismology provides the only
observational constraint, although indirect, on the
p+p $\rightarrow$ d + e$^+$ + $\nu _e$ reaction.

Incidentally, we observe that this analysis completely
excludes the value $\Sppzero =2.9 S_{pp,SSM}$, recently
found in Ref. \cite{Ober}, by using a too rough description of
the low energy pp scattering data, see also Ref. \cite{KBpp}.

\section{Future prospects and further applications}
\label{future}

The helioseismic observation of the Sun will continue and 
the accuracy of the frequency data will improve. For solar 
physicists most hope is connected with the possibility of 
studying changes during the coming maximum of the magnetic 
activity. Astronomers interesed in the angular evolution are 
looking forward to new data which should tell us whether indeed 
solar core rotates slower than the outer layers. This is so 
strange that most of them hesitate to believe it now.

We do not expect revelations as far as internal structure is 
concerned. The chances to find something that would shake the 
standard model of stellar evolution seem very small. 
Unquestionably, however, there is a room for improvement and 
helioseismology may help.

As a few examples 
of relevance to calculation of precise solar models for evaluation
of the neutrino fluxes as well to modeling evolution of other stars,
we  briefly outline a few topics we are presently working on.


\begin{figure}[tb]
\vspace{-1.3cm}
\epsfig{file=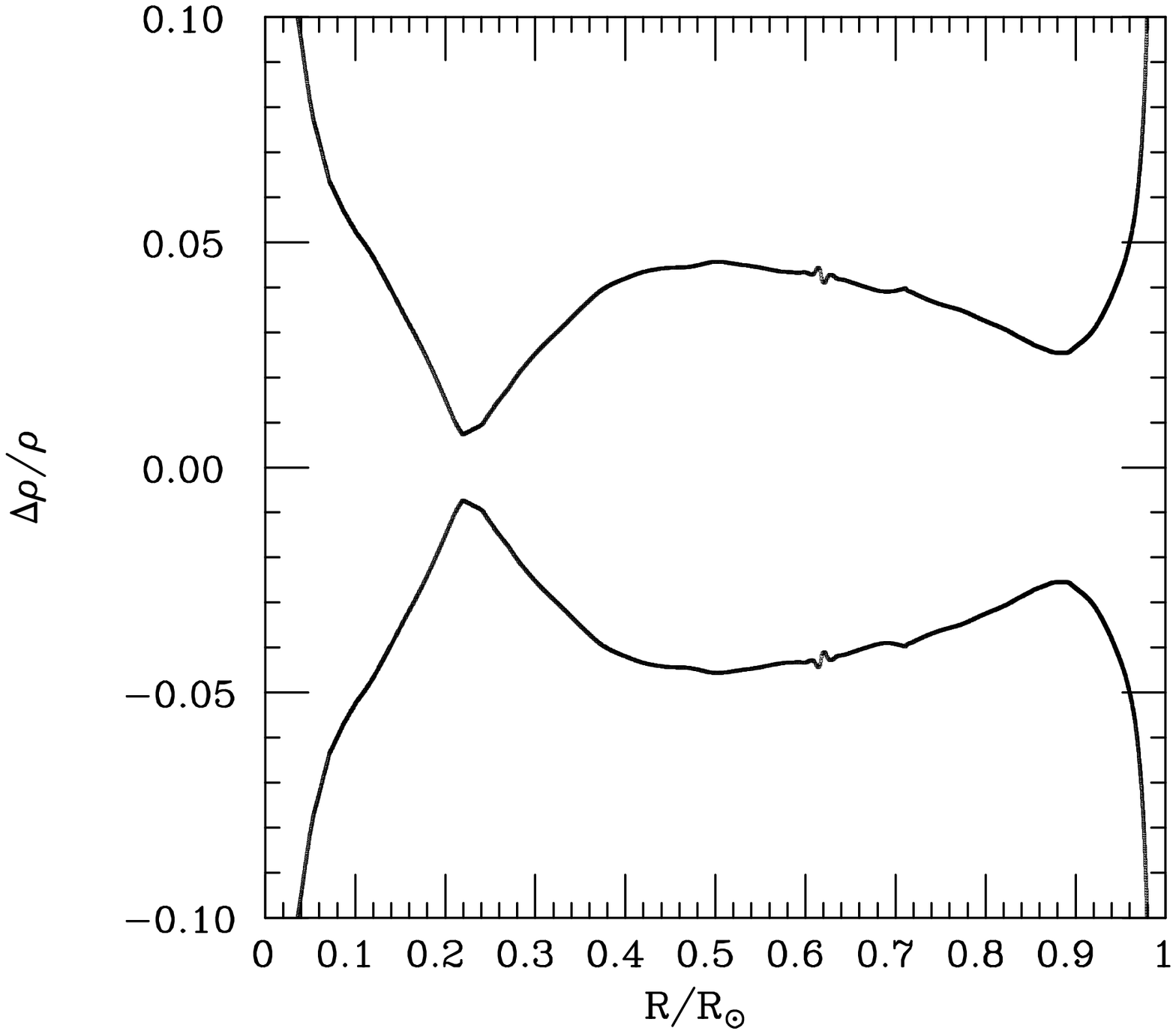,
width=0.9\hsize,angle=90}
\vspace{-2.1cm}
\caption[den]{
The estimated global relative uncertainty on $\rho$.
             }
 \label{figrho}
\end{figure}

\subsection{The solar density profile}

From the knowledge of  $U=P/\rho$  along the solar profile,
by using the hydrostatic equilibrium equation, one can extract
the solar density (see \eg~ \cite{rhow}). 
We performed a systematic analysis of the
uncertainties so as to provide a quantitative estimate of
the accuracy $\Delta \rho /\rho$, the preliminary result
being presented in Fig. \ref{figrho}.

We remark that accurate predictions of the solar density,
are important for determining the $\nu_e \rightarrow \nu_x$
transition probability in neutrino oscillation models
where neutrino conversion is due to matter effects, 
see \eg~\cite{denfluct,valle,burgers}.

\subsection{Solar opacity}

As remarked above, solar opacity $\kappa(R)$ is the more uncertain
{\em ingredient} of solar models. By using the additional information
provided by helioseismology one can try  to transform it
into an {\em output} of frequencies measurements. 
The idea is as follows: as we know $U=P/\rho$, we  have one more 
information on solar structure, and we can use one of
the equations of stellar equilibrium/evolution for getting
$\kappa$. 

In fact, in the region beyond the energy producing core 
and below the convective envelope ($0.2<R/R_\odot<0.65$) one can
use the radiative transport equation  in the form:
\begin{equation}
\label{kappa1}
 \kappa (R) \rho(R)= \frac{ -16 \pi \sigma} {3 L_\odot}
\frac {dT^4}{dR} R^2
\end{equation}
where $L_\odot$ is the solar luminosity. In the same region
the perfect gas equation  (for fully ionized H and He and
for neutral Z) holds to high accuracy and the (almost) uniform 
abundance of He and Z can be inferred from the observed properties
of the solar photosphera. In this way temperature gradients 
can be expressed in terms of gradients of $U$ (again determined
from helioseismology) so that 
the photon mean free path, $\lambda=1/(\kappa \rho)$,
can be extracted, possibly
with a few per cent accuracy.
 
This could provide an observational, although indirect, test of opacity
calculation for plasma condition as in the solar interior, see also
Ref. \cite{opahelios}.

\subsection{The statistical distribution of nuclei in the solar core}

Already two decades ago, Kacharov \etal~\cite{Kacharov} 
and Clayton \etal~\cite{Clayton},
speculated  that the high energy tail of the distribution of
protons in the solar core could depart from the Maxwell
exponential form,
\begin{equation}
\label{eqmax}
\frac{dN_{Max}(E)}{dE}  \rightarrow \frac{dN_{Max}}{dE}   exp[-\delta (E/KT) ^2]
\end{equation}
 as an attempt to reduce 
the prediction of $^8$B neutrino flux by invoking a depletion
in the number of protons with energy high enough  to be 
captured by $^7$Be nuclei.

Recently interest in this matter was revived by Quarati \etal~ 
\cite{Quarati}, who provided some argument for such a depletion,
in the frame of non extensive Tsallis statistics.

In fact, sub barrier nuclear reaction provide a good laboratory
for testing the high energy tail of the distribution, as they
involve particles near the  Gamow peak $E_G$, which is generally
larger than the average thermal energy $KT$. 
Tiny deviation from the Maxwellian distribution,
undetectable near $KT$, would be amplified
at higher energies, see Eq. \ref{eqmax}.

If the energy distribution deviates from the maxwellian
form, the reaction rates $\langle \sigma v \rangle$ would differ from
those calculated with the Maxwell distribution

Clearly this also holds for the basic 
 p+p$\rightarrow$ e$^+$+$\nu_e$
reaction. As we discussed in sect. \ref{secspp}, its rate can
be constrained by helioseismology. In the same way, possible deviations
from the Maxwell distribution can be studied, see Ref. \cite{Lissia}

~\\
~\\

\thanks{We are   grateful to Venja Berezinsky for a
friendly collaboration during the last few years.
We thank the
director of Osservatorio Astronomico di Collurania, Teramo, and 
the director of Laboratori Nazionali del Gran Sasso for their 
hospitality.}

\end{document}